\documentclass{elsart}
\usepackage{graphicx}
\usepackage{amsmath, amssymb}

\begin{document}

\begin{frontmatter}

\title{Collision of two breathers at surface of deep water}

\author[NGU,Landau]{A.I. Dyachenko,\corauthref{cor}}
\ead{alexd@landau.ac.ru}
\corauth[cor]{Corresponding author.}
\author[NGU,Tucson,Landau,Phian]{V.E. Zakharov}
{
\ead{zakharov@math.arizona.edu}
}
\author[NGU]{and D.I. Kachulin}
\address[NGU]{Novosibirsk State University, Pirogova 2,Novosibirsk-90, 630090, Russia}
\address[Tucson]{Department of Mathematics, University of
Arizona, Tucson, AZ, 857201, USA}
\address[Phian]{Physical Institute of RAS,
Leninskiy prospekt, 53, Moscow, 119991, Russia}
\address[Landau]{Landau Institute for Theoretical Physics,
2 Kosygin str., Moscow, 119334, Russia}

\begin{abstract}

We applied canonical transformation to water wave equation not only to remove cubic nonlinear terms but to simplify drastically fourth order terms in Hamiltonian. This transformation explicitly uses the fact of vanishing exact four waves interaction for water gravity waves for 2D potential fluid. After the transformation well-known but cumbersome Zakharov equation is drastically simplified and can be written in X-space in compact way. This new equation is very suitable as for analytic study as for numerical simulation. Localized in space breather-type solution was found. Numerical simulation of collision of two such breathers strongly supports hypothesis of integrability of 2-D free surface hydrodynamics.
\end{abstract}


\end{frontmatter}

\date{\today}


\section{Introduction}
The work described here is motivated by two remarkable facts regarding one-dimensional free surface hydrodynamics:
\begin{itemize}
\item In \cite{DZ94} it was shown that four-wave interaction coefficient vanishes on the resonant manifold
\begin{eqnarray}\label{RES_MAN}
k + k_1 & = & k_2 + k_3,\cr
\omega_k + \omega_{k_1} & = & \omega_{k_2} + \omega_{k_3}.\nonumber
\end{eqnarray}
 \item In \cite{DZ08,ZD10} it was demonstrated that giant breather, highly nonlinear,
exists on the fluid surface without radiation. Moreover, space-time spectrum of the breather consists of waves propagating in the same direction.
\end{itemize}
These two facts bring the idea that fourth order wave interactions can be drastically simplified by some canonical transformation of the Hamiltonian. Below we show how this transformation looks like. Dynamic equation derived using this transformation is very elegant and simple. It can be easily generalized for "almost" one-dimensional waves.

\section{Compact equation}

A one-dimensional potential flow of an ideal incompressible
fluid with a free surface in a gravity field fluid is
described by the following set of equations:
\begin{eqnarray}\nonumber
\phi_{xx} + \phi_{zz} &=& 0 \hspace{1cm} (\phi_z\to 0, z\to -\infty), \nonumber\\
\eta_t + \eta_x\phi_x &=& \phi_z\bigg|_{z=\eta}\nonumber\cr
\phi_t + \frac{1}{2}(\phi_x^2 + \phi_z^2) + g\eta &=& {0\bigg|_{z=\eta}};
\end{eqnarray}
here $\eta(x,t)$ - is the shape of a surface, $\phi(x,z,t)$ - is a potential
function of the flow and $g$ - is a gravitational acceleration.
	As was shown in\cite{Z68}, the variables $\eta(x,t)$ and
$\psi(x,t) = \phi(x,z,t)\bigg|_{z=\eta}$ are canonically conjugated,
and satisfy the equations

$$\frac{\partial \psi}{\partial t} =
-\frac{\delta H}{\delta \eta} \hspace{2cm}
  \frac{\partial \eta}{\partial t} = \frac{\delta H}{\delta \psi}.$$

\noindent Here $H=K+U$ is the total energy of the fluid with the following
kinetic and potential energy terms:

$$ K = \frac{1}{2}\int\!dx\!\int_{-\infty}^\eta\,v^2\!dz \hspace{1cm}
   U = \frac{g}{2}\int \eta^2\!dx$$

It is convenient to introduce  normal complex variable $a_k$:

\begin{equation}\nonumber
\eta_k =  \sqrt{\frac{\omega_k}{2g}}(a_k+a^*_{-k}) \hspace{.5cm}
\psi_k =  -i\sqrt{\frac{g}{2\omega_k}}(a_k-a^*_{-k})
\end{equation}

\noindent here $\omega_k = \sqrt{gk}$ -is the dispersion law for
the gravity waves, and Fourier transformations $\psi(x)\rightarrow\psi_k$ and $\eta(x)\rightarrow\eta_k$ are defined as follows:
$$
f_k =  \frac{1}{\sqrt{2\pi}}\int f(x)e^{-ikx}dx, \hspace{2em}
f(x) = \frac{1}{\sqrt{2\pi}}\int f_k e^{+ikx}dk.
$$
\noindent Hamiltonian can be expanded in an infinite series in powers of
$a_k$ (see\cite{{Z68},{CYS80}})

\begin{equation}\nonumber
H = H_2 + H_3 + H4 + \ldots
\end{equation}

This variable $a_k$ satisfies the equation

\begin{equation}\nonumber
\frac{\partial a_k}{\partial t} + i\frac{\delta H}{\delta a_k^*}=0,
\end{equation}
\noindent where

\begin{eqnarray}\nonumber
H_2 & = & \int\!\omega_k a_k a_k^*dk,\nonumber\cr
H_3 & = & \int\!V^{k}_{k_1 k_2}\{a_k^*a_{k_1}a_{k_2}+a_k a_{k_1}^*a_{k_2}^*\}
\delta_{k-k_1-k_2}\!dkdk_1dk_2\nonumber\cr
&+&\frac{1}{3}\int\!U_{k k_1 k_2}\{a_ka_{k_1}a_{k_2}+a_k^*a_{k_1}^*a_{k_2}^*\}
\delta_{k+k_1+k_2}\!dkdk_1dk_2\nonumber
\end{eqnarray}

\begin{eqnarray}\nonumber
V^{k}_{k_1 k_2} & = & \frac{g^{\frac{1}{4}}}{8\sqrt{\pi}}\left(
\left(k\over{k_1k_2}\right)^{1\over4}L_{k_1k_2}-
\left(k_2\over{kk_1}\right)^{1\over4}L_{-kk_1} -
\left(k_1\over{kk_2}\right)^{1\over4}L_{-kk_2}
\right)\nonumber\cr
U_{k k_1 k_2} & = & \frac{g^{\frac{1}{4}}}{8\sqrt{\pi}}\left(
\left(k\over{k_1k_2}\right)^{1\over4}L_{k_1k_2}+
\left(k_2\over{kk_1}\right)^{1\over4}L_{kk_1} +
\left(k_1\over{kk_2}\right)^{1\over4}L_{kk_2}
\right)
\end{eqnarray}

\begin{equation}\nonumber
L_{kk_1} = (\vec{k} \vec{k_1}) + |k||k_1|
\end{equation}
Fourth order part of Hamiltonian is the following:
\begin{eqnarray}\nonumber
H_4 &=&
\frac{1}{2}\int W_{k_1k_2}^{k_3k_4}a_{k_1}^*a_{k_2}^*a_{k_3}a_{k_4}\delta_{k_1+k_2-k_3-k_4}dk_1dk_2dk_3dk_4 +\cr
&+&\frac{1}{3}\int G_{k_1k_2k_3}^{k_4}
(a_{k_1}^*a_{k_2}^*a_{k_3}^*a_{k_4}+a_{k_1}a_{k_2}a_{k_3}a_{k_4}^*\delta_{k_1+k_2+k_3-k_4}dk_1dk_2dk_3dk_4 +\cr
&+&\frac{1}{12}\int R_{k_1k_2k_3k_4}
(a_{k_1}^*a_{k_2}^*a_{k_3}^*a_{k_4}^*+a_{k_1}a_{k_2}a_{k_3}a_{k_4})\delta_{k_1+k_2+k_3+k_4}dk_1dk_2dk_3dk_4
\end{eqnarray}
Here $W_{k_1k_2}^{k_3k_4}$, $G_{k_1k_2k_3}^{k_4}$ and $R_{k_1k_2k_3k_4}$ are equal to:
\begin{eqnarray}\nonumber
W_{k_1k_2}^{k_3k_4} = \frac{-1}{32\pi}&&\left [
M_{-k_3-k_4}^{k_1k_2} + M_{k_1k_2}^{-k_3-k_4} - M_{k_2-k_4}^{k_1-k_3} -
M_{k_1-k_4}^{k_2-k_3} - M_{k_2-k_3}^{k_1-k_4} - M_{k_1-k_3}^{k_2-k_4}
\right ]\cr
G_{k_1k_2k_3}^{k_4} = \frac{-1}{32\pi}&&\left [
M_{k_1k_2}^{k_3-k_4} + M_{k_1k_3}^{k_2-k_4} + M_{k_2k_3}^{k_1-k_4} -
M_{k_3-k_4}^{k_1k_2} - M_{k_2-k_4}^{k_1k_3} - M_{k_1-k_4}^{k_2k_3}
\right ]\cr
R_{k_1k_2k_3k_4} = \frac{-1}{32\pi}&&\left [
M_{k_1k_2}^{k_3k_4} + M_{k_1k_3}^{k_2k_4} + M_{k_1k_4}^{k_2k_3} +
M_{k_2k_3}^{k_1k_4} + M_{k_2k_4}^{k_1k_3} + M_{k_3k_4}^{k_1k_2}
\right ].
\end{eqnarray}
Here
$$
M_{k_1k_2}^{k_3k_4} = |k_1k_2|^{\frac{3}{4}}|k_3k_4|^{\frac{1}{4}}(|k_1+k_3|+|k_1+k_4|+|k_2+k_3|+|k_2+k_4|-2|k_1|-2|k_2|).
$$
\noindent Now one can apply canonical transformation from variables $a_k$ to $b_k$ to exclude non resonant cubic terms along with non resonant fourth order terms with coefficients $G_{k_1k_2k_3}^{k_4}$ and $R_{k_1k_2k_3k_4}$. This transformation up to the accuracy $O(b^5)$ has the form \cite{Z68, KRS91,ZLF92}:
\begin{eqnarray}\label{TRANSFORMATION}
a_k &=& b_k+\int\Gamma^{k}_{k_1k_2}b_{k_1}b_{k_2}\delta_{k-k_1-k_2}dk_1dk_2
-2\int\Gamma^{k_2}_{kk_1}b_{k_1}^*b_{k_2}\delta_{k+k_1-k_2}dk_1dk_2+\cr
&+&\int\Gamma_{kk_1k_2}b_{k_1}^*b_{k_2}^*\delta_{k+k_1+k_2}dk_1dk_2
+\int B_{kk_1}^{k_2k_3}b_{k_1}^*b_{k_2}b_{k_3}\delta_{k+k_1-k_2-k_3}dk_1dk_2dk_3+\cr
&+&\int C_{kk_1k_2}^{k_3}b_{k_1}^*b_{k_2}^*b_{k_3}\delta_{k+k_1+k_2-k_3}dk_1dk_2dk_3
+\int S_{kk_1k_2k_3}b_{k_1}^*b_{k_2}^*b_{k_3}^*\delta_{k+k_1+k_2+k_3}dk_1dk_2dk_3.
\end{eqnarray}
Here
\begin{eqnarray}\nonumber
B_{kk_1}^{k_2k_3} &=& \Gamma^{k_1}_{k_1-k_2} \Gamma^{k_3}_{kk_3-k}
+ \Gamma^{k_1}_{k_3k_1-k_3} \Gamma^{k_2}_{kk_2-k} -
 \Gamma^{k}_{k_2k-k_2} \Gamma^{k_3}_{k_1k_3-k_1}
- \Gamma^{k_1}_{k_3k_1-k_3} \Gamma^{k_2}_{k_1k_2-k_1}-\nonumber\cr
&-& \Gamma^{k+k_1}_{kk_1} \Gamma^{k_2+k_3}_{k_2k_3}
+ \Gamma_{-k-k_1kk_1} \Gamma_{-k_2-k_3k_2k_3}+\tilde B_{k_2k_3}^{kk_1},\nonumber
\end{eqnarray}

\begin{eqnarray}\nonumber
\Gamma^{k}_{k_1k_2} = -{V^{k}_{k_1k_2}\over{\omega_k-\omega_{k_1}-\omega_{k_2}}},\hspace{0.8cm}
\Gamma_{kk_1k_2} = -{U_{kk_1k_2}\over{\omega_k+\omega_{k_1}+\omega_{k_2}}}
\nonumber
\end{eqnarray}
and $\tilde B_{k_2k_3}^{kk_1}$ is an arbitrary function satisfying the following symmetry conditions:
\begin{eqnarray}\nonumber
\tilde B_{k_2k_3}^{kk_1} = \tilde B_{k_2k_3}^{k_1k} =
\tilde B_{k_3k_2}^{kk_1} = -(\tilde B_{kk_1}^{k_2k_3})^*.
\end{eqnarray}
\noindent Coefficients $C_{kk_1k_2}^{k_3}$ and $S_{kk_1k_2k_3}$ provide vanishing corresponding forth-order terms in the new Hamiltonian.

Details of this transformation can be found in \cite{DZ2011-1,DZ2011-2}. In K-space Hamiltonian has the form:
\begin{eqnarray}\nonumber
{\cal H}_ = \int \omega_k|b_k|^2 + \frac{1}{2}\int \tilde T_{k_1k_2}^{k_3k_4}b_{k_1}^*b_{k_2}^*b_{k_3}b_{k_4}\delta_{k_1+k_2-k_3-k_4}dk_1dk_2dk_3dk_4 
\end{eqnarray}
In X-space it corresponds to:
\begin{eqnarray}\label{SPACE}
{\cal H} =  \int\!b^*\hat\omega_k bdx +
\frac{i}{16}\int\!\left [ {b^*}^2 \frac{\partial}{\partial x} ({b'}^2) -
b^2 \frac{\partial}{\partial x} ({{b^*}'}^2) \right ]dx
-\frac{1}{4}\int\!|b|^2 \cdot \hat K(|b'|^2)dx.
\end{eqnarray}
Here $b' = \frac{\partial }{\partial x}b$.
After integrating by parts Hamiltonian acquires very nice form:
\begin{eqnarray}\label{SPACE_NICE}
{\cal H} =  \int\!b^*\hat\omega_k bdx +
\frac{1}{4}\int\!|b'|^2\left [\frac{i}{2}(bb'^* - b^*b') -\hat K|b|^2 \right ] dx.
\end{eqnarray}

Corresponding equation of motion is the following:
\begin{eqnarray}\label{MotionSPACE}
i\frac{\partial b}{\partial t} = \hat\omega_k b
+\frac{i}{8}\left [ b^* \frac{\partial}{\partial x} ({b'}^2) -
\frac{\partial}{\partial x}( {b^*}' \frac{\partial}{\partial x}b^2) \right ]
-\frac{1}{4} \left [ b \cdot \hat K(|b'|^2) - \frac{\partial}{\partial x}(b'\hat K (|b|^2))\right ].
\end{eqnarray}

\section{Monochromatic wave and modulational instability}

Monochromatic wave
\begin{equation}\label{Mono}
b(x) = B_0 e^{i(k_0x - \omega_0 t)}
\end{equation}
is the simplest solution of (\ref{MotionSPACE}). Indeed, plugging (\ref{Mono}) in  to the equation (\ref{MotionSPACE}) one can get the following relation
\begin{equation}\label{StokesShift}
\omega_0 = \omega_{k_0} +\frac{1}{2}k_0^3 |B_0|^2.
\end{equation}
Recalling transformation from $a_k$ to $b_k$ one can see that for waves with small amplitude ( $a_k\simeq b_k$)
$$
|B_0|^2 = \frac{\omega_{k_0}}{k_0}\eta_0^2,
$$
and relation (\ref{StokesShift}) coincides with well known Stokes correction to the frequency due to finite wave amplitude.
\begin{equation}\label{SS}
\omega_0 = \omega_{k_0}(1 +\frac{1}{2}k_0^2 |\eta_0|^2).
\end{equation}

\subsection{Modulational instability of monochromatic wave}

Let us consider perturbation to the solution
$$
b = B_{0}e^{i(k_0x - \omega_0 t)}
$$
where
$$
B_0 = \frac{1}{\sqrt{2\pi}}\int b_{k_0}e^{i(k_0x - kx)}dx
$$
and
$$
\omega_0 = \omega_{k_0} + \frac{1}{2}|B_0|^2 k_0^3, \hspace{2em} \frac{1}{4\pi}|b_{k_0}|^2 k_0^3 =
\frac{1}{2}T_{k_0 k_0}^{k_0 k_0}|b_{k_0}|^2.
$$
Perturbed solution has the following form:
\begin{equation}\label{PERTK}
b \Rightarrow (b_{k_0} +
\delta b_{k_0+k}e^{-i\Omega_k t} + \delta b_{k_0-k}e^{-i\Omega_{-k} t})e^{-i\omega_0 t}.
\end{equation}
with the following condition:
$$
\Omega_{k} = -\Omega_{-k}
$$
Plugging perturbed solution (\ref{PERTK}) in to the equation
$$
i\dot b_k = \omega_k b_k +\frac{1}{2}\int \tilde T_{k_2k_3}^{kk_1}b^*_{k_1}b_{k_2}b_{k_3}
\delta_{k+k_1-k_2-k_3}\!dk_1\!dk_2\!dk_3
$$
we get the sum of two independent equations:
\begin{eqnarray}\nonumber
\hspace{-1cm}\left [
i\delta \dot b_{k_0+k} +(\omega_0+\Omega_{k})\delta b_{k_0+k}\right .
&=& \left . \omega_{k_0+k}\delta b_{k_0+k}
+\tilde T^{k_0+k k_0}_{k_0+k k_0}|b_{k_0}|^2\delta b_{k_0+k}+
\frac{1}{2}\tilde T_{k_0 k_0}^{k_0+k k_0-k}b_{k_0}^2\delta b^*_{k_0-k}
\right ] e^{-i\omega_0t - i\Omega_{k} t}+\cr
\hspace{-1cm}+\left [
i\delta \dot b_{k_0-k} +(\omega_0+\Omega_{-k})\delta b_{k_0-k} \right .
&=& \left . \omega_{k_0-k}\delta b_{k_0-k}
+\tilde T_{k_0-k k_0}^{k_0-k k_0}|b_{k_0}|^2\delta b_{k_0-k}+
\frac{1}{2}\tilde T_{k_0 k_0}^{k_0-k k_0+k}b_{k_0}^2\delta b^*_{k_0+k}
\right ]
e^{-i\omega_0t - i\Omega_{-k} t}.
\end{eqnarray}
Expressions for $\tilde T^{k_0+k k_0}_{k_0+k k_0}$ and $T_{k_0 k_0}^{k_0-k k_0+k}$ can be easily obtained:
\begin{eqnarray}\label{TT}
\tilde T^{k_0+k k_0}_{k_0+k k_0} &=& \frac{k_0^3}{4\pi} + \frac{k_0(3k_0-|k|)}{4\pi}k +
\frac{k_0}{4\pi}(k_0^2 - k_0|k| +k^2),\\
T_{k_0 k_0}^{k_0-k k_0+k} &=& \frac{k_0}{2\pi}(k_0^2 - k_0|k| - \frac{k^2}{2}).
\end{eqnarray}
Looking at even and odd powers of $k$ one can see that
$$
\Omega_k = \frac{\omega_{k_0+k} - \omega_{k_0-k}}{2} + \frac{|B_0|^2}{2}(3k_0-|k|)k
$$
Let us denote
$$
d(k) = \frac{\omega_{k_0+k} - 2\omega_{k_0}+\omega_{k_0-k}}{2}.
$$
Then
$$
i\delta \dot b_{k_0+k} = d(k)\delta b_{k_0+k} + \frac{|B_0|^2k_0}{2}(k_0^2-k_0|k|+k^2)\delta b_{k_0+k}+
\frac{B_0^2k_0}{2}(k_0^2-k_0|k|-\frac{k^2}{2})\delta b^*_{k_0-k}.
$$
Suppose $\delta b_{k_0+k}$ growth as
$$
\delta b_{k_0+k} \Rightarrow \delta b_{k_0+k}e^{\gamma_k t}
$$
one can easily obtain the following formula for $\gamma_k$:
\begin{equation}\label{GrowtRate}
\gamma_k^2 = \left [-d(k)-\frac{3|B_0|^2}{4}k_0 k^2\right]\left [d(k) +|B_0|^2 k_0(k_0-\frac{|k|}{2})^2\right ].
\end{equation}
If we introduce steepness of the carrier wave $\omega_{k_0}\mu^2 = |B_0|^2k_0^2$ and approximate $d(k)$ as
$$
d(k) \simeq -\frac{1}{8}\omega''_{k_0}k^2 = -\frac{1}{8}\omega_{k_0}\frac{k^2}{k_0^2},
$$
then for growth rate is
\begin{equation}\label{GR}
\gamma_k^2 = \frac{1}{8}\frac{\omega_{k_0}^2}{k_0^4}(1-{\bf 6\mu^2})k^2
\left [\mu^2(k_0 - {\bf \frac{|k|}{2}})^2-\frac{k^2}{8} \right ].
\end{equation}
The difference between this formula and well-known expression derived from the nonlinear Schrodinger equation is highlighted by two boldfaced terms.

\section{Breathers and numerical sumulation of its collisions}

Breather is the localized solution of (\ref{MotionSPACE}) of the following type:
\begin{equation}\label{NLS}
b(x,t) = B(x-Vt) e^{i(k_0x - \omega_0 t)},
\end{equation}
where $k_0$ is the wavenumber of the carrier wave, $V$ is the group velocity and $\omega_0$ is the frequency close to $\omega_{k_0}$.
In the Fourier space breather can be written as follow:
\begin{equation}\label{BK}
b_k(t) = e^{-i(\Omega t + Vk)}\phi_k,
\end{equation}
where $\Omega$ is close to $\frac{\omega_{k_0}}{2}$.

For $\phi_k$ the following equation is valid:
\begin{equation}\label{BREATHER}
(\Omega + Vk -\omega_k) \phi_k = \int \tilde T_{k k_1,k_2 k_3}\phi_{k_1}^* \phi_{k_2}\phi_{k_3}
\delta_{k+k_1-k_2-k_3}dk_1dk_2dk_3.
\end{equation}
One can treat $\phi_k$ as pure real function of $k$.

\begin{figure}
\includegraphics[angle=-00,width=14.0cm]{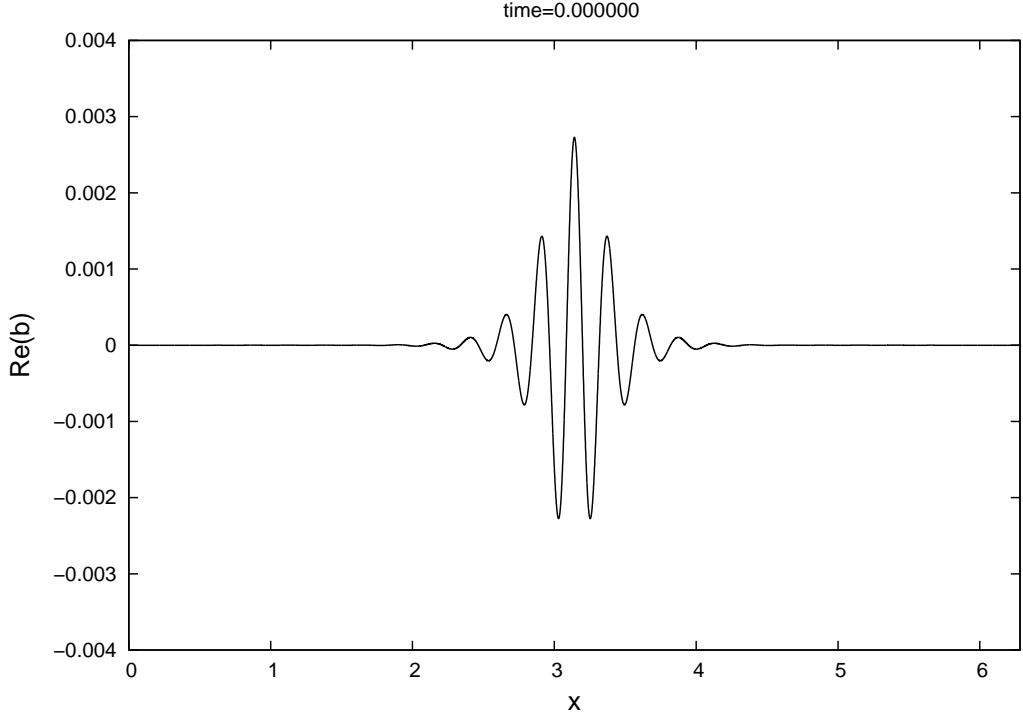}%
\caption{Real part of $b(x)$ with $V=0.1$ and $\Omega=2.53$.  \label{FIG_01}}
\end{figure}

\begin{figure}
\includegraphics[angle=-00,width=14.0cm]{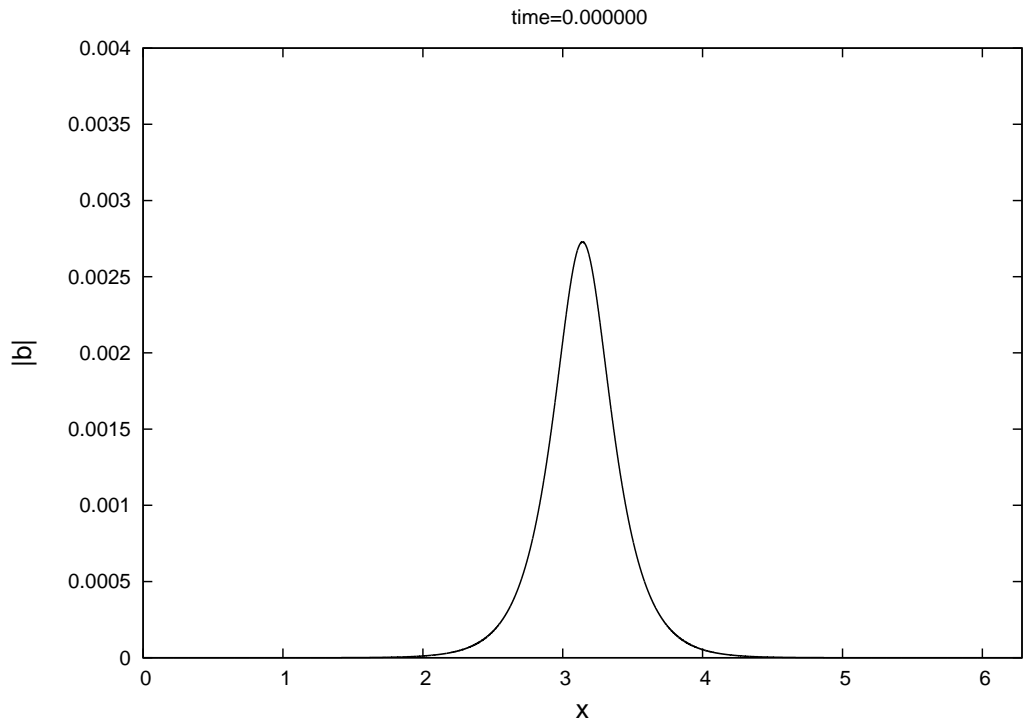}%
\caption{Modulus of $b(x)$ with $V=0.1$ and $\Omega=2.53$. (Recall envelope in the Nonlinear Shrodinger Equation \label{FIG_02}}
\end{figure}

\begin{figure}
\includegraphics[angle=-00,width=14.0cm]{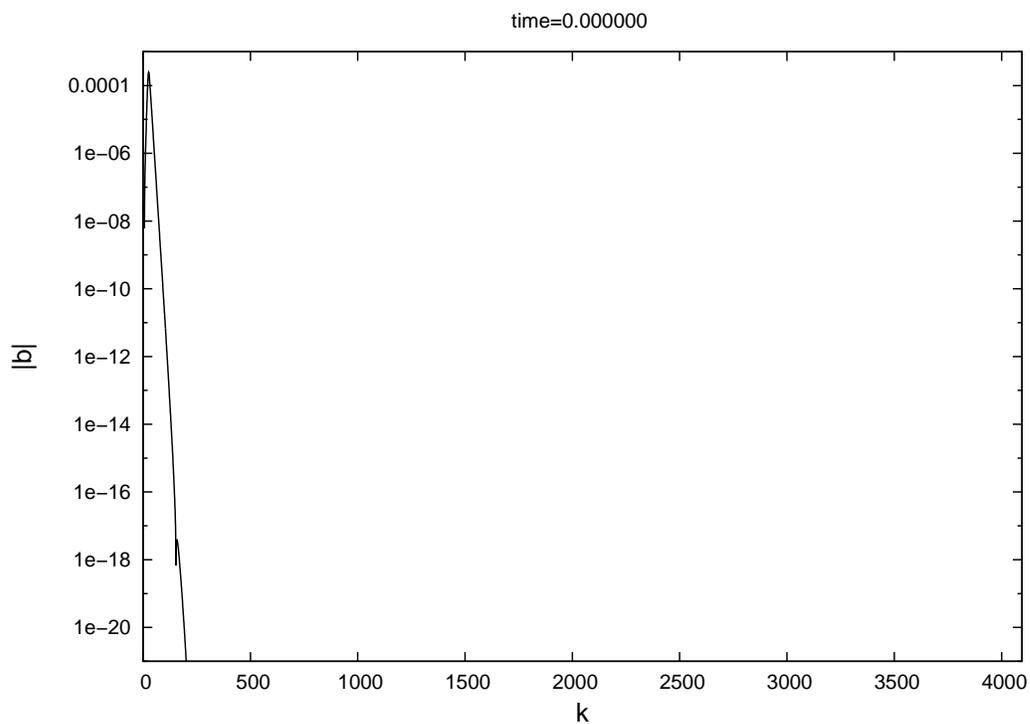}%
\caption{Spectrum of $b(x)$ with $V=0.1$ and $\Omega=2.53$.  \label{FIG_03}}
\end{figure}

To solve equation (\ref{BREATHER}) one can use Petviashvili iteration method:
\begin{equation}\label{PETV}
(\Omega + Vk -\omega_k) \phi_k^{n+1} = M^n\int \tilde T_{k k_1,k_2 k_3}{\phi_{k_1}^*}^n \phi_{k_2}^n\phi_{k_3}^n
\delta_{k+k_1-k_2-k_3}dk_1dk_2dk_3.
\end{equation}
Petviashvili coefficient $M$ is the following:
$$
M^n = \left [ \frac{<\phi_k^n (\Omega + Vk -\omega_k)\phi_k^n>}{<\phi_k^n\int \tilde T_{k k_1,k_2 k_3}{\phi_{k_1}^*}^n \phi_{k_2}^n\phi_{k_3}^n
\delta_{k+k_1-k_2-k_3}dk_1dk_2dk_3 >}\right ] ^{\frac{3}{2}}.
$$
Below we present typical numerical solution of (\ref{BREATHER}). Calculation were made in the periodic domain $2\pi$ with carier wavenumber $k_0\sim 25$, $V=0.1$ and $\Omega=2.53$.\footnote{steepness of carrier wave $\mu$ must not exeed the crical value $\sqrt{\frac{1}{6}}$ as it follows from formula for growt rate of modulatinal instability (\ref{GR})}.
In the Figures \ref{FIG_01}, \ref{FIG_02}, \ref{FIG_03} one can see real part of$b(x)$, modulus of $b(x)$ and Fourier spectrum of $b(x)$.

\begin{figure}
\includegraphics[angle=-00,width=14.0cm]{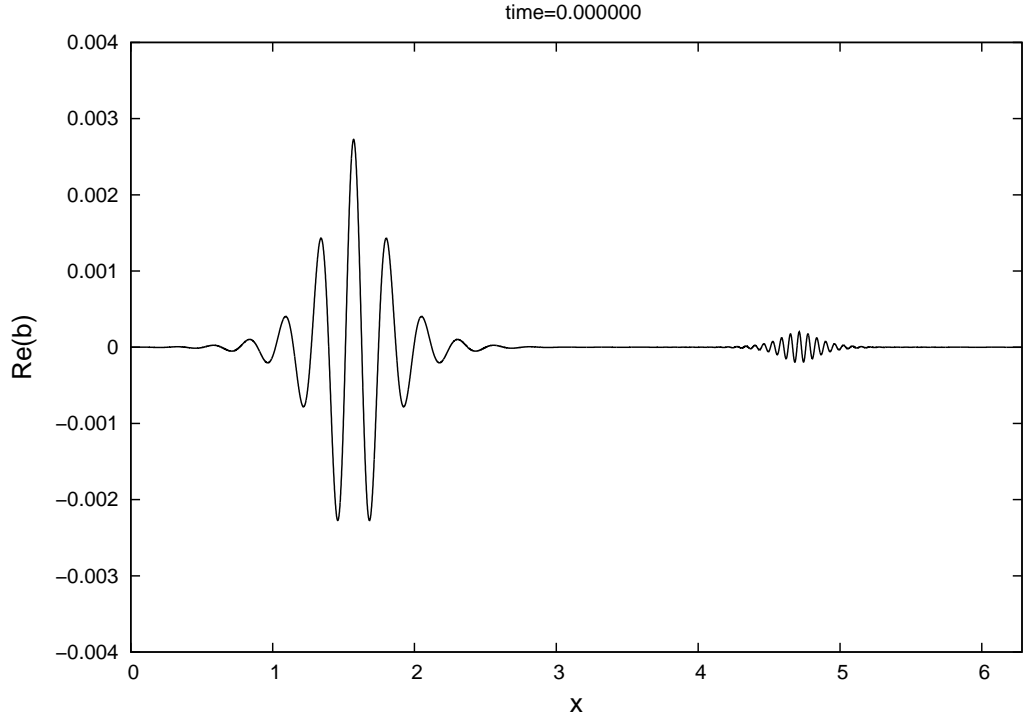}%
\caption{Initial condition with two breathers.  \label{FIG_04}}
\end{figure}

\begin{figure}
\includegraphics[angle=-00,width=14.0cm]{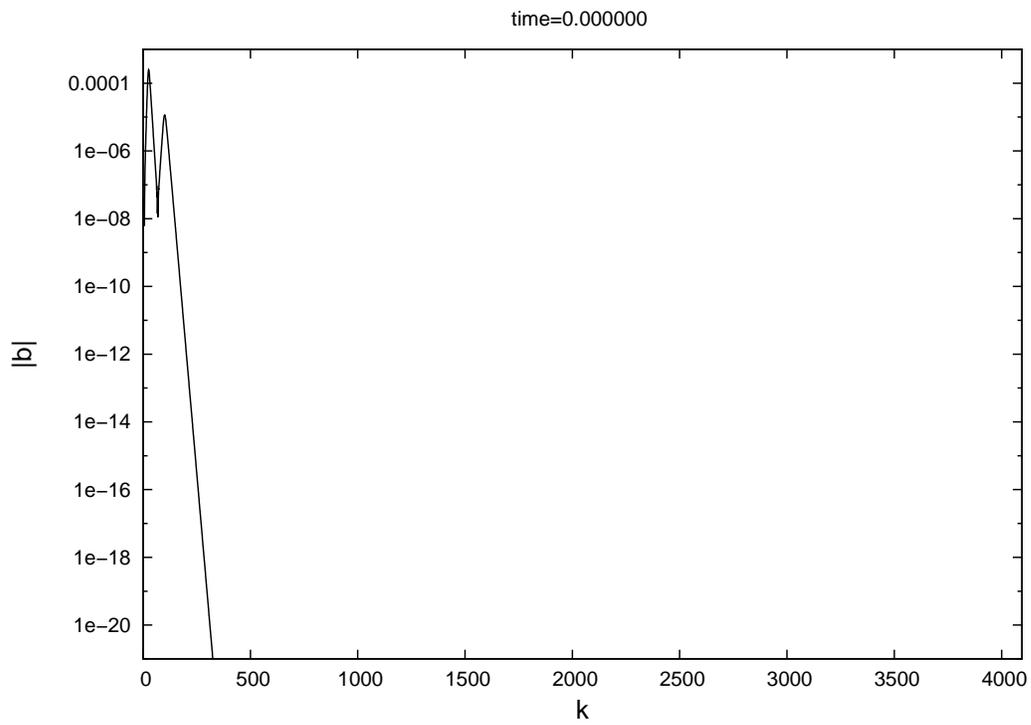}%
\caption{Fourier spectrum of two breathers.  \label{FIG_05}}
\end{figure}

\begin{figure}
\includegraphics[angle=-00,width=14.0cm]{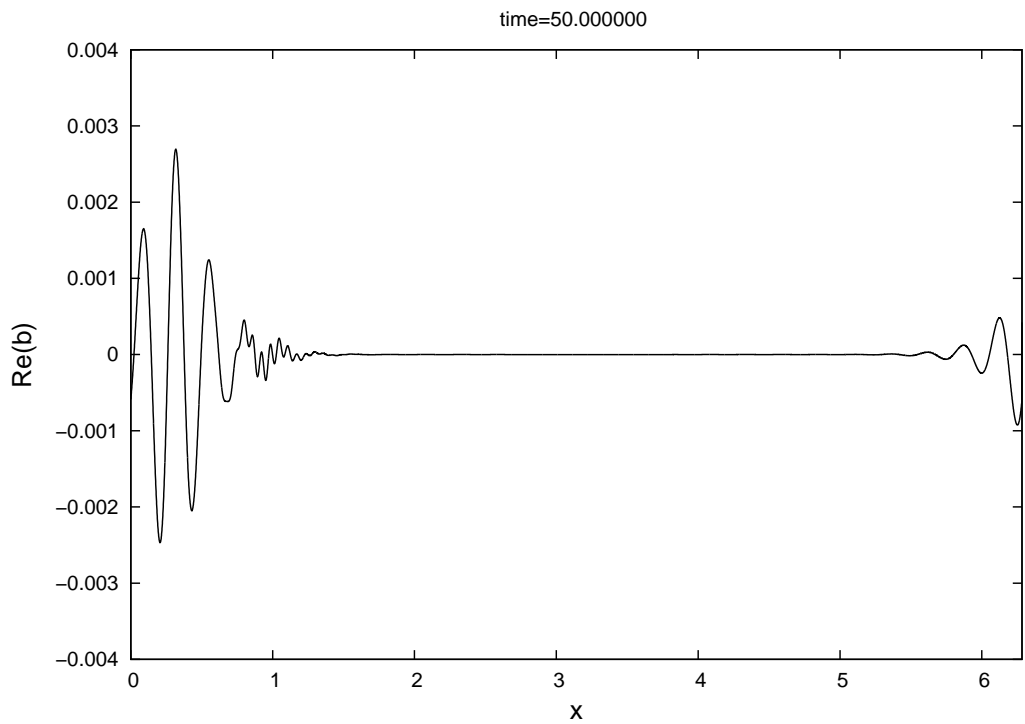}%
\caption{Two breathers just before collision.  \label{FIG_06}}
\end{figure}

\begin{figure}
\includegraphics[angle=-00,width=14.0cm]{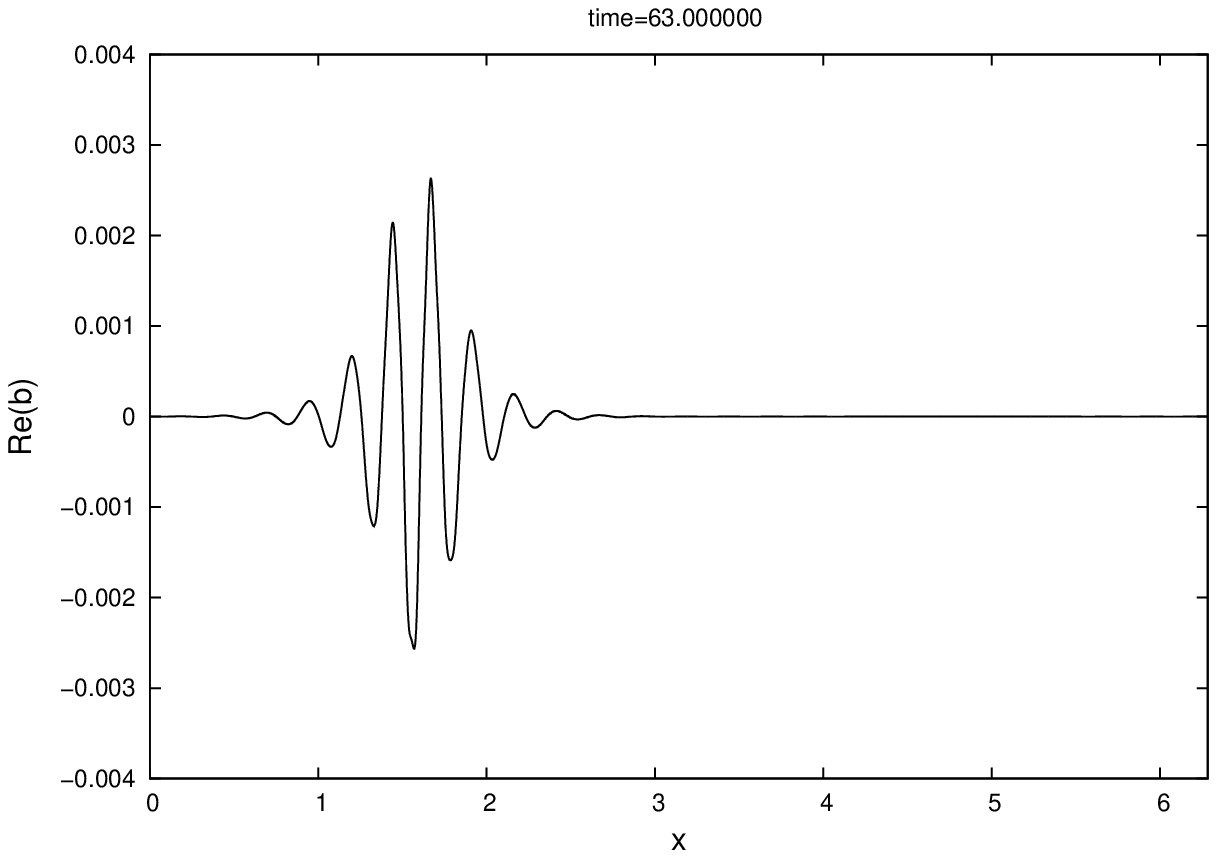}%
\caption{Two breathers collides.  \label{FIG_07}}
\end{figure}

\begin{figure}
\includegraphics[angle=-00,width=14.0cm]{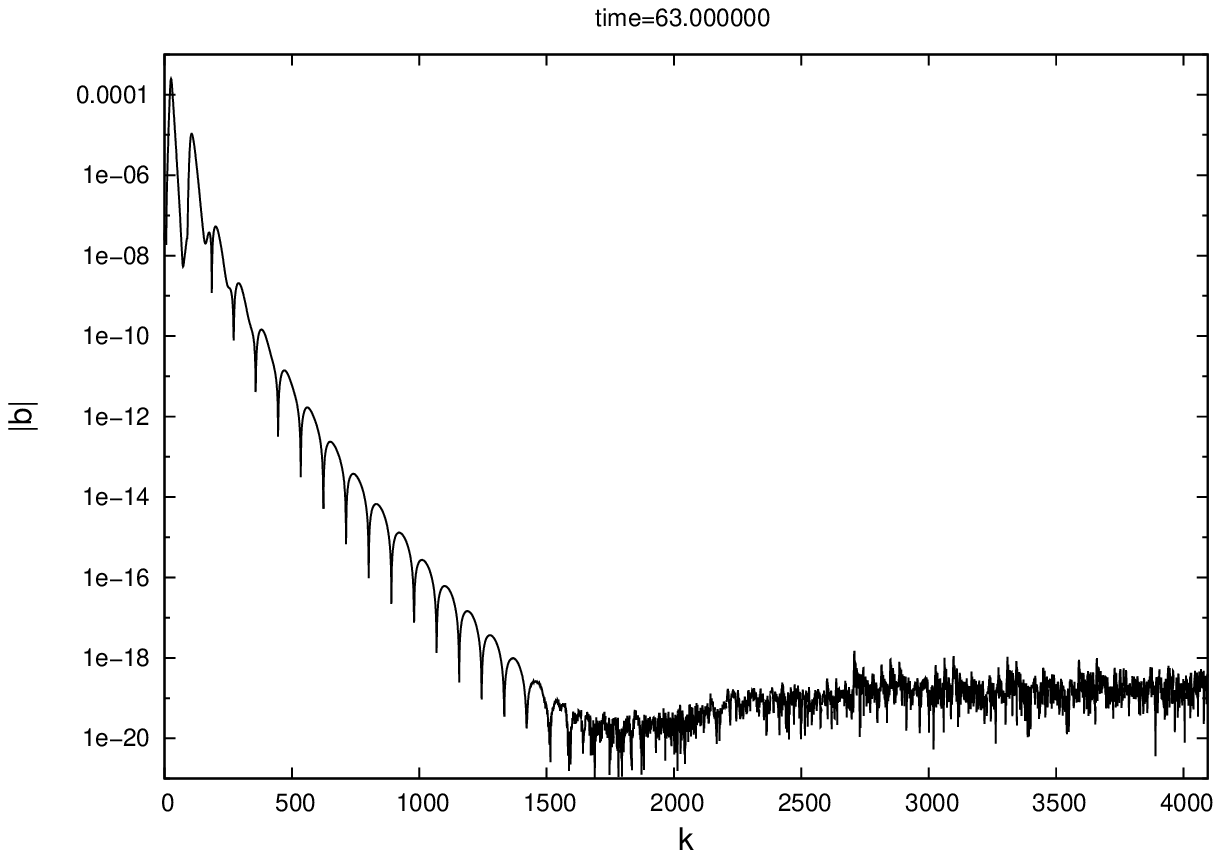}%
\caption{Fourier spectrum at the moment of collision.  \label{FIG_08}}
\end{figure}

\begin{figure}
\includegraphics[angle=-00,width=14.0cm]{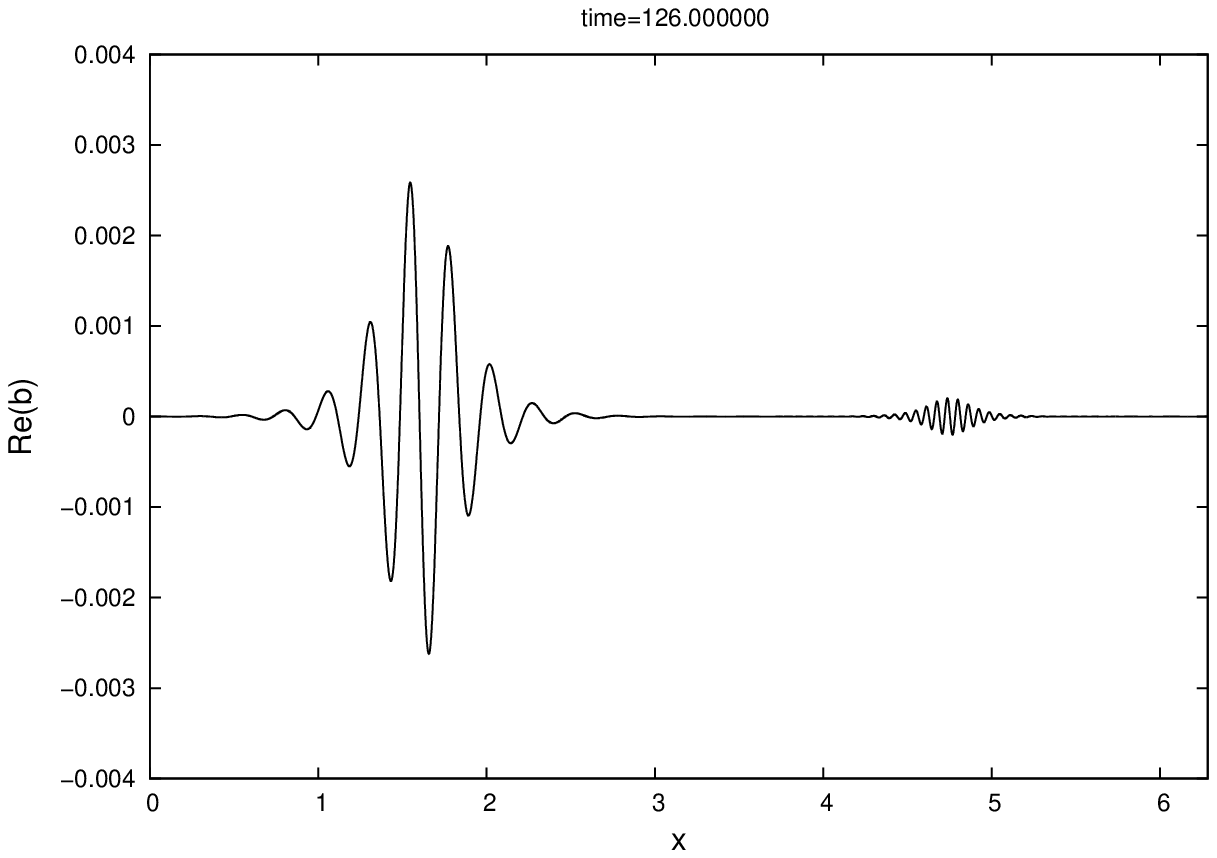}%
\caption{Two breathers after collision.  \label{FIG_09}}
\end{figure}

\begin{figure}
\includegraphics[angle=-00,width=14.0cm]{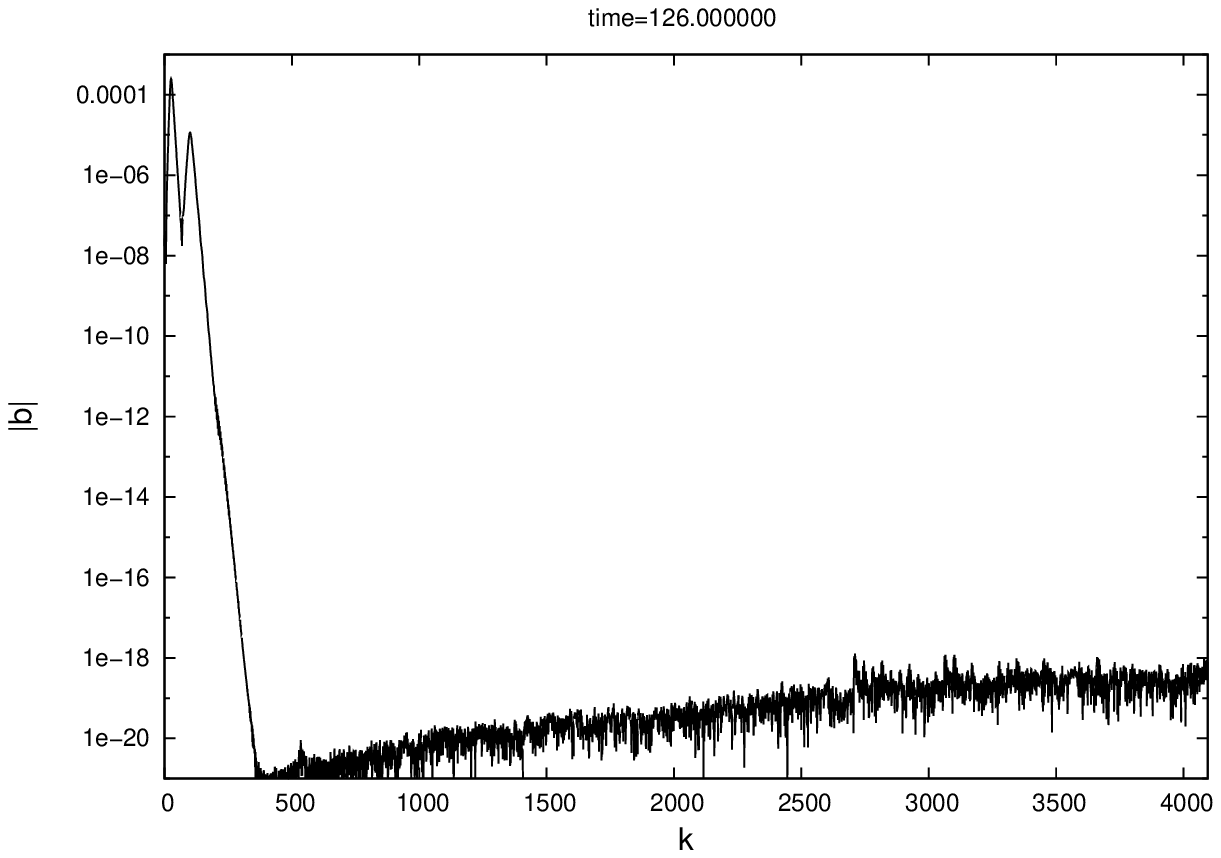}%
\caption{Fourier spectrum at the moment of collision.  \label{FIG_10}}
\end{figure}

Very important question from the point of view of integrability of the equation (\ref{MotionSPACE}) is the question about collision of two breathers. To study breathers collision we performed the following numerical simulation:
\begin{itemize}
\item As initial condition we have used two beathers separated in space (distance was $\pi$.)
\item First breather has the following parameters: $\Omega_1 = 5.1$, $V_1=0.05$. Carrier wave number appears to be $\sim 100$.
\item For the second breather - $\Omega_2 = 2.53$, $V_2=0.1$. Carrier wave number appears to be $\sim 25$
\end{itemize}
This initial condition is show in Figure \ref{FIG_04}. 
Its Fourier spectrum is shown in Figure \ref{FIG_05}.
After time $\frac{\pi}{(V_2-V_1)}\simeq 62.8$ breathers collide. In the Figures \ref{FIG_06} and \ref{FIG_07} one can see breathers at the time close to collision ($t = 50$) and at the moment of collision ($t = 63$).
Fourier spectrum of two breathers at $t=63$ is shown in Figure \ref{FIG_08}.
And finally we show the picture of two breathers at $t=126$ when they separated again at distance $\simeq\pi$. Real part of $b(x)$ and Fourier spectrum of that is given in Figures \ref{FIG_09}, \ref{FIG_10}.
So, the simulation demonstrates no interaction of breathers. It suggests that equation (\ref{MotionSPACE}) is integrable.
\section{Conclusion}

Simple equation describing evolution of 1-D water waves is derived. Derivation of this equation is based on the important property of vanishing four-wave interaction for gravity water waves. This property allows to simplify drastically well-known Zakharov's equation for water waves, which is very cumbersome. Written in X-space instead of K-space, it allows further analytical and numerical study. The equation has breather-type solution which was found numerically using Petviashvili iteration method. Numerical simulation of collision of two breathers shows behavior which is typical for integrable system. It can be considered as numerical proof of integrability. 

 This new equation can be generalized for the "almost" 2-D waves, or "almost" 3-D fluid. When considering waves slightly inhomogeneous in transverse direction, one can think in the spirit of Kadomtsev-Petviashvili equation for Korteveg-de-Vries equation, namely one can treat now frequency $\omega_k$ as two dimensional, $\omega_{k_x,ky}$, while leaving coefficient $\tilde T_{k_2k_3}^{kk_1}$  not dependent on $y$. $b$ now depends on both $x$ and $y$:
\begin{eqnarray}\label{KP}
{\cal H} =  \int\!b^*\hat\omega_{k_x, k_y} bdxdy +
\frac{1}{4}\int\!|b'_x|^2\left [\frac{i}{2}(bb'^*_x - b^*b'_x) -\hat K_x|b|^2 \right ] dxdy.
\end{eqnarray}

\section{Acknowledgments}

This work was supported by Grant of Government of Russian Federation for support of scientific research, carried under direction of leading scientists in russian universities N11.G34.31.0035 (leading scientist -- Zakharov V.E., GOU VPO ``Novosibirsk State University'').
Also was it was supported by the US Army Corps of Engineers Grant W912-BU-08-P-0143, by ONR Grant N00014-10-1-0991, NSF Grant DMS 0404577, Grant NOPP "TSA-a two scale approximation for wind-generated ocean surface waves", RFBR Grant 09-01-00631 and RFBR Grant 09-05-13605, the Program "Fundamental Problems in Nonlinear Dynamics"  from the RAS Presidium, and Grant "Leading Scientific Schools of Russia".

\end{document}